     \documentstyle[proceedings,numreferences]{crckapb} 

\input{psfig}

\newcommand{\ket}[1]{\left|#1\right\rangle}
\newcommand{\bra}[1]{\left\langle#1\right|}
\def\Hcontr{{\cal H}_{\rm ctrl}}
\def\P{{\cal P}}
\newcommand{\color}[1]{}

\begin{opening}
\title{Dephasing and Renormalization\protect\\ 
in Quantum Two-Level Systems}

\author{Alexander Shnirman}
\institute{Institut f\"ur 
Theoretische Festk\"orperphysik, Universit\"at Karlsruhe,  
76128 Karlsruhe, Germany}
\author{Gerd Sch\"on}
\institute{Institut f\"ur 
Theoretische Festk\"orperphysik, Universit\"at Karlsruhe,  
76128 Karlsruhe, Germany; and 
Forschungszentrum Karlsruhe, 
Institut f\"ur Nanotechnologie, 
76021 Karlsruhe, Germany}

\end{opening}

\runningtitle{Dephasing and Renormalization ...}

\begin{document}


\begin{abstract}

Motivated by fundamental questions about the loss of 
phase coherence at low temperature we consider relaxation, dephasing
and renormalization effects in quantum two-level systems which are
coupled to a dissipative environment. We observe that experimental 
conditions, e.g., details of the initial state preparation, 
determine to which extent the environment leads to dephasing 
or to renormalization effects. We analyze an exactly 
solvable limit where the relation between both
can be demonstrated explicitly. We also study the 
effects of dephasing and renormalization on response functions. 

\end{abstract}

\section{Introduction}

The dynamics of quantum two-level systems has always
been at the focus of interest, but recently attracted increased
attention because of the prospects of quantum state engineering and
related low-temperature experiments. A crucial requirement for many of
these concepts is the preservation of phase coherence in the presence of
a noisy environment. Typically one lacks a detailed microscopic
description of the noise source, but frequently it is sufficient to model
the environment by a bath of harmonic oscillators, with frequency
spectrum adjusted to reproduce the observed power spectrum.  The
resulting `spin-boson' models have been studied in the literature, in
particular the one with bilinear coupling and `Ohmic' spectrum, but
spin-boson models with different power spectra appear equally
interesting in view of several experiments.  In this article
we, therefore, analyze spin-boson model with general power spectra
with respect to relaxation, dephasing and renormalization
processes at low temperatures. We study the dephasing of nonequilibrium
initial states. We show 
how the state preparation affects the effective high-frequency cut-off and
separates renormalization from dephasing effects. We also 
demonstrate how dephasing processes influence low-temperature linear
response and correlation functions.

\section{Spin-boson model}
\label{Sec:spin-boson}

In this section we review the theory and properties of spin-boson
models, which have been studied extensively before (see the reviews
\cite{LeggettRMP,WeissBook}).  
A quantum two-level system is modeled by a spin degree of freedom in a
magnetic field. It is coupled linearly to an oscillator bath
representing the environment. The total Hamiltonian is
\begin{equation}
{\cal H}=\Hcontr + \sigma_z \sum_j c_j (a^{\phantom
 \dagger}_j+a^\dagger_j) + {\cal H}_{\rm b} \; ,
\label{Eq:SpinBoson}
\end{equation}
where the controlled part is
$ \Hcontr = -\frac{1}{2}B_z\;\sigma_z- \frac{1}{2}B_x\;\sigma_x =
-\frac{1}{2} \Delta E \,(\cos\theta\;\sigma_z+\sin\theta\;\sigma_x)$,
the oscillator bath is described by
$
{\cal H}_{\rm b}= \sum_j \hbar \omega_j \, a^\dagger_ja^{\phantom
\dagger}_j\, ,
$
and the bath `force' operator
$X = \sum_j c_j (a^{\phantom \dagger}_j+a^\dagger_j)$ is assumed to
couple linearly to $\sigma_z$.
For later convenience  $\Hcontr$ has also 
been written in terms of a mixing angle 
$\theta = \tan^{-1} (B_x/B_z)$, depending on the direction of
the magnetic field, and the energy splitting
of the eigenstates, $\Delta E = \sqrt{B_x^2+B_z^2}$. 

In thermal equilibrium the Fourier transform of the symmetrized
correlation function of the force operator is given by
\begin{equation}
\label{Eq:X-J}
S_X(\omega)
\equiv \left\langle [ X(t), X(t') ]_+ \right\rangle_\omega
= 2 \hbar J(\omega) \coth \frac{\hbar \, \omega}{2 k_{\rm B}T} \;.
\end{equation}
Here the bath spectral density has been introduced, defined by
$
J(\omega) \equiv {\pi\over\hbar}\sum_j c_j^2 \;
\delta(\omega-\omega_j) \,.
$
At low frequencies it typically has a power-law form
up to a high-frequency cut-off $\omega_c$,
\begin{equation}
J(\omega) = \frac{\pi}{2}\,\hbar \alpha\, \omega_0^{1-s}\omega^s
\Theta(\omega_c - \omega)
\ .
\label{J}
\end{equation}
Generally one distinguishes Ohmic ($s=1$), sub-Ohmic ($s<1$),
and super-Ohmic  ($s>1$) spectra.
In Eq.~(\ref{J}) an additional frequency scale $\omega_0$ has been introduced.
Since it appears only in the combination $\alpha\omega_0^{1-s}$
it is arbitrary and in Ref.~\cite{LeggettRMP} has been chosen equal
to the high-frequency cut-off. Here 
we prefer to distinguish both frequencies, since the cut-off
$\omega_c$ will play an important role in what follows.  

The spin-boson model has been studied
mostly for the case where an Ohmic bath is coupled
linearly to the spin. One reason is that linear damping,
proportional to the velocity, is encountered frequently in
real systems. Another is that suitable systems with Ohmic
damping show a quantum phase transition at a critical strength 
of the dissipation, $\alpha_{\rm cr} \sim 1$. 
On the other hand, in the context of quantum-state engineering we 
should concentrate on systems with weak damping,
but allow for general spectra of the fluctuations.

Spin-boson models with sub-Ohmic damping ($0<s<1$) have been 
considered earlier~\cite{LeggettRMP,WeissBook}
but did not attract much attention. It was argued that 
sub-Ohmic dissipation would totally suppress coherence, 
transitions between the states of the
two-level system would happen only at
finite temperatures and would be incoherent. 
At zero temperature the system should be localized in one 
of the eigenstates of $\sigma_z$, since
the bath renormalizes the off-diagonal part of the Hamiltonian 
$B_x$ to zero. This scenario is correct for intermediate
to strong damping. It is, however, not correct for weak damping. Indeed the
`NIBA' approximation developed in Ref.~\cite{LeggettRMP}
fails in the weak-coupling limit for transverse noise, while a more 
accurate renormalization procedure~\cite{Kehrein_Mielke_PhysLettA96} 
predicts damped coherent behavior. In the context of quantum-state engineering 
we are interested in precisely this {\it coherent sub-Ohmic} regime. 
We will demonstrate a simple criterion which allows to define a border 
between coherent and incoherent regimes.

A further reason to study the sub-Ohmic case is that it allows us to 
mimic the universally observed $1/f$ noise.
For instance, a bath with $s=0$ and $J(\omega) = (\pi/2) \alpha
\hbar \omega_0$ produces at low frequencies,
$\hbar \omega\ll k_{\rm B}T$, a $1/f$ noise spectrum
$
S_X(\omega) = E_{1/f}^2/|\omega|
$
with $E_{1/f}^2=2\pi\alpha\hbar\omega_0 k_{\rm B}T$.
Since frequently nonequilibrium sources are responsible for the $1/f$ noise, 
the temperature here should be regarded as a fitting parameter rather
than a thermodynamic quantity. 

Below we will also consider the super-Ohmic case ($s>1$)
as it allows us to study renormalization effects clearly.

\section{Relaxation and dephasing}

We first consider the Ohmic case in the weak damping regime, $\alpha \ll 1$.  
Still we distinguish two regimes: the `Hamiltonian-dominated' regime,
which is realized when $\Delta E$ is large enough, and the
`noise-dominated' regime, which is realized, e.g., at degeneracy points
where $\Delta E \to 0$. The exact border between both regimes will 
be specified below.

In the Hamiltonian-dominated regime it is natural to describe the evolution 
of the system in the eigenstates of $\Hcontr$, which are
\begin{eqnarray}
\label{Eigen_Basis}
        \ket{0} = \;\;\; \cos{\theta\over2} \ket{\uparrow} +
\sin{\theta\over2} \ket{\downarrow} \;\; \mbox{and} \;\; \ket{1} &=& -
\sin{\theta\over2} \ket{\uparrow} + \cos{\theta\over2}
\ket{\downarrow} \ .
\end{eqnarray}
Denoting by $\tau_x$ and $\tau_z$ the Pauli matrices in the eigenbasis,
we have
\begin{equation}
\label{Eq:Spin_Boson_Eigen_Basis}
{\cal H} = -{1\over 2}\Delta E\, \tau_z + (\sin\theta\;\tau_x +
\cos\theta\;\tau_z) \; X + {\cal H_{\rm b}} \ .
\end{equation} 
\vspace{3mm}

Two different time scales describe the evolution in the
spin-boson model.  The first, the
dephasing time scale $\tau_\varphi$, characterizes the decay of the
off-diagonal elements of the qubit's reduced density matrix
$\hat\rho(t)$ in the preferred eigenbasis (\ref{Eigen_Basis}), or,
equivalently  of the expectation values of the operators
$\tau_{\pm} \equiv (1/2) (\tau_x \pm i \tau_y)$. 
Frequently dephasing processes lead to an exponential 
long-time dependence,
\begin{equation}
\label{Eq:Rho_pm_dephasing} 
\langle \tau_{\pm}(t) \rangle 
\equiv \mbox{tr}\,[\tau_{\pm} \hat\rho(t) ]
\propto \langle \tau_{\pm}(0) \rangle\; e^{\mp
i\Delta E t/\hbar}\; e^{-t/\tau_\varphi} \ ,
\end{equation}
but other decay laws occur as well and will be discussed below. 
The second, the relaxation time scale $\tau_{\rm relax}$, characterizes how
diagonal entries tend to their  equilibrium values, 
\begin{equation}
\label{Eq:Rho_z_relaxation}
\langle \tau_z(t) \rangle - \langle \tau_z (\infty)\rangle \propto
e^{-t/\tau_{\rm relax}} \, ,
\end{equation}
where $\langle \tau_z (\infty)\rangle = \tanh(\Delta E/2k_{\rm B}T)$.

In Refs.~\cite{LeggettRMP,WeissBook} the dephasing and relaxation
times 
were evaluated in a path-integral
technique.  In the regime $\alpha \ll 1$ it is easier to
employ the perturbative (diagrammatic) technique developed in 
Ref.~\cite{Schoeller_PRB} and the standard Bloch-Redfield approximation. 
The rates are
\begin{eqnarray}
&&\Gamma_{\rm relax} \equiv \tau_{\rm relax}^{-1}
=
\frac{1}{\hbar^2} \sin^2\theta \; S_X\left(\omega =
\Delta E/\hbar\right)
\label{Eq:relaxation}
\; , \\ 
&&\Gamma_\varphi \equiv \tau_\varphi^{-1}
=
\frac{1}{2}\;\Gamma_{\rm relax} + \frac{1}{\hbar^2} \cos^2\theta \, 
S_X(\omega = 0)
\label{Eq:dephasing}
\; ,
\end{eqnarray}
where $S_X(\omega) = 
\pi\alpha \hbar^2\omega 
\coth(\hbar\omega/2k_{\rm B}T)$. 
We observe that only the `transverse' part of the fluctuating field
$X$, coupling to $\tau_x$ and proportional to $\sin\theta$,  
induces transitions between the
eigenstates (\ref{Eigen_Basis}) of the unperturbed system.  Thus the
relaxation rate\footnote{The equilibration is due to two processes,
excitation $\ket{0}\to\ket{1}$ and relaxation 
$\ket{1}\to\ket{0}$, with rates
$\Gamma_{+/-} \propto
\langle X(t)X(t')\rangle_{\omega = \pm\Delta E/\hbar}$.
Both rates are related by a detailed balance condition, and
the equilibrium value $\langle \tau_z (\infty)\rangle$ depends on both.
On the other hand, $\Gamma_{\rm relax}$ is determined by the 
sum of both rates, i.e., the symmetrized noise 
power spectrum $S_X$.} (\ref{Eq:relaxation}) is proportional to
$\sin^2\theta$.  
The `longitudinal' part of the coupling of $X$ to $\tau_z$, which is
proportional to $\cos\theta$, does not induce relaxation processes.
It does, however, contribute to dephasing since 
it leads to fluctuations of the eigenenergies and, thus, to a
random relative phase between the two eigenstates.  
This is the origin of the `pure' dephasing contribution to 
Eq.~(\ref{Eq:dephasing}), 
which is proportional to $\cos^2\theta$. 
We rewrite Eq.~(\ref{Eq:dephasing}) as 
$\Gamma_\varphi = \frac{1}{2} \Gamma_{\rm relax} + 
\cos^2\theta\,\Gamma^*_\varphi$, where 
$\Gamma^*_\varphi = S_X(\omega=0)/\hbar^2 = 2\pi\alpha k_{\rm B} T/\hbar$
is the pure dephasing rate. 

The pure dephasing rate $\Gamma^*_\varphi$ characterizes the strength
of the dissipative  
part of the Hamiltonian, while $\Delta E$ characterizes the coherent part.
The Hamiltonian-dominated regime is realized when $\Delta E \gg
\Gamma^*_\varphi$,  
while the noise-dominated regime is realized in the opposite case.

In the noise-dominated regime, $\Delta E \ll \Gamma^*_\varphi$,
the coupling to the bath is the dominant part of the total Hamiltonian.
Therefore, it is more convenient to discuss the problem in the
eigenbasis of the observable $\sigma_z$ to which the bath is
coupled. The spin can tunnel incoherently between the two eigenstates
of $\sigma_z$.  One can again employ the perturbative analysis
\cite{Schoeller_PRB}, but use directly the Markov instead of the
Bloch-Redfield approximation.  The resulting rates are given by
\begin{eqnarray}
\Gamma_{\rm relax}  &=& B_x^2/\Gamma^*_\varphi =
B_x^2 /(2\pi\hbar\alpha k_{\rm B}T) 
\,\\
\nonumber
\Gamma_{\rm \varphi} &=& \Gamma^*_\varphi = 2\pi\hbar\alpha k_{\rm B}T\ .
\label{Eq:Tau_Deph_Zeno}
\end{eqnarray}  
In this regime the dephasing is much faster than the relaxation.  In
fact, as a function of the coupling strength $\alpha$ the dephasing and
relaxation rates evolve in opposite directions.  
The $\alpha$-dependence of the relaxation rate is an indication of the Zeno 
(watchdog) effect~\cite{Harris_Stodolsky}: the environment frequently
`observes' the state of the spin, thus preventing it from tunneling.

\section{Longitudinal coupling, exact solution for factorized 
initial conditions}
\label{sec:long_exact}

The last forms of Eqs.~(\ref{Eq:relaxation}) and (\ref{Eq:dephasing}) 
express the two rates in terms of the noise power spectrum at the
relevant frequencies. These are the
level spacing of the two-state system and zero frequency, respectively. 
The expressions apply in the weak-coupling limit for 
spectra which are regular at these frequencies. For the relaxation rate 
(\ref{Eq:relaxation}) the generalization to sub- and 
super-Ohmic cases merely requires  substituting the relevant
$S_X(\omega=\Delta E/\hbar)$. However, this does not work for the
pure dephasing contribution. Indeed $S_X(\omega=0)$ is infinite
in the sub-Ohmic regime, while it vanishes for the super-Ohmic 
case and the Ohmic case at $T=0$.
As we will see this does not imply
infinitely fast or slow dephasing in these cases. To analyze these cases we
study the exactly soluble model of longitudinal coupling, $\theta = 0$.

Dephasing processes are contained in the time evolution of the quantity 
$\langle \tau_+(t)\rangle$ obtained after tracing out the bath. This quantity 
can be evaluated analytically for $\theta=0$ 
(when $\tau_+ = \sigma_+$) for an initial state which is described 
by a factorized 
density matrix $\hat\rho(t=0) = \hat\rho_{\rm spin} \otimes \hat\rho_{\rm 
bath}$ (which implies that the two-level system and bath are
initially disentangled).
In this case the Hamiltonian 
${\cal H} = -{1\over 2}\Delta E\, \sigma_z +
\sigma_z  X + {\cal H_{\rm b}}
$ is diagonalized by a unitary transformation 
by $U \equiv \exp\left(-i \sigma_z \Phi/2 \right)$
where the bath operator $\Phi$ is defined as
\begin{equation}
\Phi \equiv i \sum_j \frac{2c_j}{\hbar\omega_j}
(a^{\dagger}_j-a^{\phantom \dagger}_j)
\,.\end{equation}
The `polaron transformation' yields $\tilde {\cal H} = U{\cal H}U^{-1}
= -{1\over 2}\Delta E\, \sigma_z + {\cal H_{\rm b}}$.
It has a clear physical meaning: the oscillators are shifted 
in a direction depending on the state of the spin.
Next we observe that the operator $\sigma_+$ is transformed 
as $
\tilde \sigma_+ = U\sigma_+ U^{-1} = e^{-i\Phi}\sigma_+
\ ,$ and the observable of interest can be expressed as 
\begin{equation}
\langle \sigma_+(t) \rangle = {\rm Tr}\left[\hat\rho(t=0)\sigma_+(t)\right]
= 
{\rm Tr}\left[U\hat\rho(t=0)U^{-1} e^{-i\Phi(t)}\sigma_+\right]
\ .
\end{equation}
The 
time evolution of $\Phi(t)=e^{i{\cal H}_{\rm b}t} \Phi e^{-i{\cal
H}_{\rm b}t}$ is governed by the bare bath Hamiltonian. 
After some algebra, using the fact that the initial density 
matrix is factorized, we  obtain 
$\langle \sigma_+(t) \rangle \equiv \P(t)\; e^{-i\Delta E t}\;
\langle \sigma_+(0) \rangle$ where
\begin{equation}
\label{Eq:P(t) factorized}
\P(t)= {\rm Tr}\left[e^{i\Phi(0)/2} e^{- i\Phi(t)} e^{i\Phi(0)/2} 
\hat\rho_{\rm bath}\right]
\,.
\end{equation}

The expression~(\ref{Eq:P(t) factorized}) applies for any initial
state of the bath as long as it is factorized from the spin. In 
particular, we can assume that the spin was initially (for $t\le0$) kept in the 
state $\ket{\uparrow}$ and the bath had relaxed to the thermal equilibrium 
distribution for this spin value: $\hat\rho_{\rm bath} = \hat\rho_{\uparrow}
\equiv Z_{\uparrow}^{-1} e^{-\beta {\cal H}_{\uparrow}}$, where
${\cal H}_{\uparrow} = {\cal H}_{\rm b}+\sum_j c_j 
(a^{\phantom\dagger}_j+a^\dagger_j)$. 
In this case we can rewrite the density matrix as
$\hat\rho_{\rm bath} =e^{i\Phi/2}\hat\rho_{\rm b} e^{-i\Phi/2}$,
where the density matrix of the decoupled bath is given by
$\hat\rho_{\rm b}\equiv Z_{\rm b}^{-1}e^{-\beta {\cal H}_{\rm b}}$,
and the function $\P(t)$ reduces to
\begin{equation}
\label{Eq:P(t)}
\P(t) \rightarrow P(t) \equiv {\rm
Tr}\,\left(e^{-i\Phi(t)}\,e^{i\Phi}\,\hat\rho_{\rm b} \right) \; .
\end{equation}
This expression (with Fourier transform $P(E)$) has been studied
extensively in the  literature
\cite{LeggettRMP,P(E)_Panyukov_Zaikin,P(E)_Odintsov,P(E)_Nazarov,P(E)_Devoret}.
It can be expressed as $P(t)=\exp K(t)$, where
\begin{eqnarray}
\label{K(t)}
        K(t) = {4\over \pi\hbar} \int_0^{\infty} d\omega \,
        {J(\omega)\over\omega^2} \left[\coth\left({\hbar\omega\over2
        k_B T}\right)(\cos\omega t-1) -i\sin\omega t\right] \ .
\end{eqnarray}
For an Ohmic bath at non-zero temperature
and not too short times, $t>\hbar/k_{\rm B}T$, it reduces to
$ {\rm Re} \, K(t) \approx -S_X(\omega=0) t/\hbar^2 =
-2\pi\,\alpha\,k_{\rm B}T\, t /\hbar \, ,$
consistent with Eq.~(\ref{Eq:dephasing}) in the limit $\theta = 0$.

For $1/\omega_c<t<\hbar/k_{\rm B}T$, and thus for all times at $T=0$,
one still finds a decay 
of $\langle \sigma_+(t)\rangle$ governed 
by ${\rm Re}\;K(t) \approx -2\;\alpha\;\ln(\omega_c t)$, which implies
a power-law decay 
\begin{equation}
\label{Eq:power_law_dephasing}
\langle\sigma_+(t)\rangle = (\omega_c t)^{-2\alpha} e^{- i\Delta E t/\hbar}
\langle\sigma_+(0)\rangle
\ .
\end{equation}
Thus even at $T=0$, when $S_X(\omega=0)=0$, the off-diagonal elements 
of the density matrix decay in time. All oscillators up 
to the high-frequency cut-off $\omega_c$ contribute to this decay.
The physical meaning of this result will be discussed later. We can also 
define a cross-over temperature $T^*$ below which the power-law decay
dominates over the subsequent exponential one. A criterion is that at
$t=\hbar/k_{\rm B}T^*$ the short-time power-law decay has reduced the
off-diagonal components already substantially, i.e   
$\langle \sigma_+(\hbar/k_{\rm B}T^*) \rangle = 1/e$. This happens at
the temperature 
$k_{\rm }T^*=\hbar \omega_c \exp(-1/2\alpha)$. 
Thus the dephasing rate is $\Gamma^*_\varphi = k_{\rm B}T^*/\hbar$ for 
$T<T^*$ and $\Gamma^*_\varphi = 2\pi \alpha k_{\rm B} T/\hbar$ for
$\alpha T > T^*$.

For a sub-Ohmic bath with $0<s<1$ due to the high density of low-frequency 
oscillators exponential dephasing  
is observed even for short times,  $|\langle  
\sigma_+(t) \rangle| \propto \exp[-\alpha (\omega_0 t)^{1-s}]$
for $t<\hbar/k_{\rm B}T$, as well as 
for longer times, $|\langle 
\sigma_+(t) \rangle| \propto 
\exp[-\alpha \,Tt\,(\omega_0 t)^{1-s}]$
for $t>\hbar/k_{\rm B}T$. In the exponents 
of the short- and long-time decay laws we have omitted factors 
which are of order one, except if $s$ is close to either $1$ or 
$0$, in which case a more careful treatment is required. 
The dephasing rates resulting from the decay laws are 
$\Gamma_\varphi^* \propto T^*=\alpha^{1/(1-s)}\omega_0$ 
(cf. Ref.~\cite{Unruh}) for $T<T^*$ and 
$\Gamma_\varphi^* \propto (\alpha T/\omega_0)^{1/(2-s)}\omega_0$ 
for $T>T^*$. Again the crossover temperature $T^*$ marks the boundary
between the regimes where either the initial, temperature-independent
decay or the subsequent decay at $t>\hbar/k_{\rm B}T$ is more important.

These results allow us to 
further clarify the question of coherent vs. incoherent 
behavior in the sub-Ohmic regime. It is known from earlier 
work~\cite{LeggettRMP,WeissBook} that the dynamics of the 
spin is incoherent even at $T=0$ if 
$\Delta E \ll \alpha^{1/(1-s)}\hbar\omega_0$ (only the transverse case 
$\theta = \pi/2$ was considered). This condition implies $\Delta E \ll
\Gamma_\varphi^*$, i.e., it marks the noise-dominated regime.
We are mostly interested in the opposite, Hamiltonian-dominated
regime, $\Delta E \gg \hbar\Gamma_\varphi^*$. The NIBA approximation 
used in Ref.~\cite{LeggettRMP} fails in this limit, while a more accurate 
RG study~\cite{Kehrein_Mielke_PhysLettA96} predicts damped coherent behavior.

Finally we discuss the super-Ohmic regime, $s>1$.  
In this case, within a short time of order $\omega_c^{-1}$,
the exponent ${\rm Re}\;K(t)$ increases to a finite value 
${\rm Re}\;K(\infty) = - \alpha(\omega_c/\omega_0)^{s-1}$ 
and then remains constant  for  $\omega_c^{-1} < t < \hbar/k_{\rm B}T$
(we again omit factors of order one and note that the limit 
$s\rightarrow 1$ requires more care). 
This implies an initial reduction of the off-diagonal element 
$|\langle \sigma_+(t) \rangle|$ followed by a saturation at $|\langle
\sigma_+(t) \rangle| 
\propto \exp[-\alpha (\omega_c/\omega_0)^{s-1}]$. For $t>\hbar/k_{\rm
B}T$ an exponential  decay develops,  $|\langle 
\sigma_+(t) \rangle| \propto \exp[-\alpha \,Tt\,(\omega_0 t)^{1-s}]$,
but only if $s<2$.
This decay is always dominant and, thus, there is no crossover in this
case, i.e.,  
$T^* =0$. For $s \ge 2$ there is almost no additional decay.
 
The results obtained for different bath spectra are summarized in 
Table~\ref{tbl:time_decay}.

\begin{table}
\caption{\label{tbl:time_decay}Decay of $|\langle\sigma_+(t)\rangle|$ 
in time for different bath spectra.}

\begin{tabular}{ll}
&\\
\noindent 
$\bullet$
Ohmic: $J(\omega) = \frac{\pi}{2}\alpha \omega \Theta(\omega_c-\omega)$:&\\

\parbox{8.5cm}{
for $\omega_c^{-1} \ll t \ll T^{-1}$
\quad
$|\langle\sigma_+(t)\rangle| \approx (\omega_c\,t)^{-2\alpha}$
\\
for $\ {\phantom{\omega_c^{-1} \ll}}t \gg T^{-1}$
\quad
$|\langle\sigma_+(t)\rangle| \approx e^{-2\pi\alpha T\,t}$
}
&
\parbox{5cm}{
\psfig{width=4cm,figure=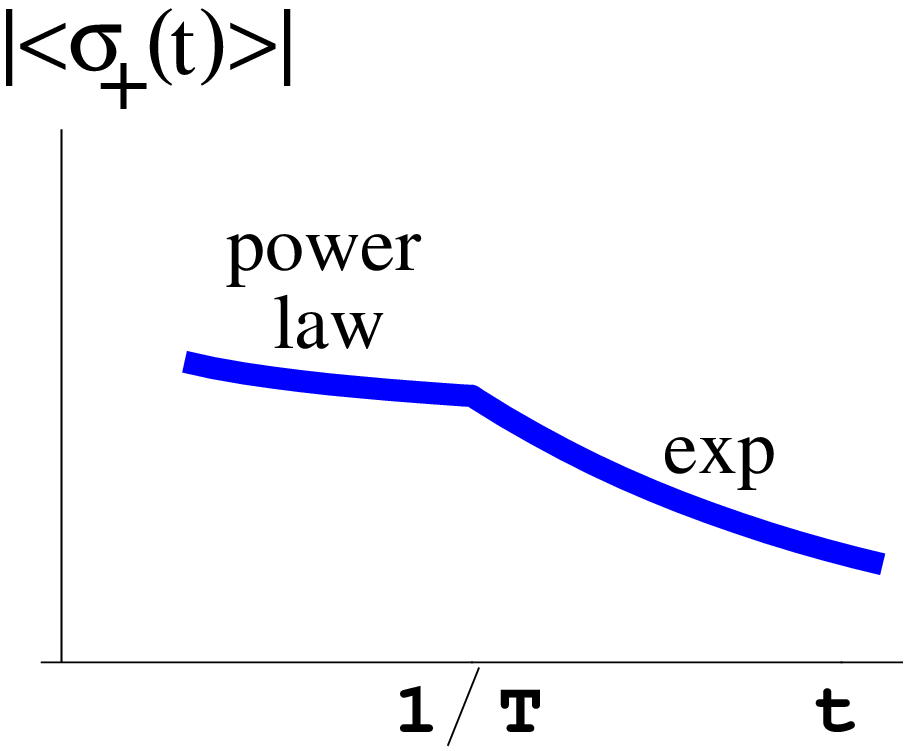}
}\\

\noindent 
$\bullet$
Sub-Ohmic: $J(\omega) = \frac{\pi}{2}\alpha
\omega_0^{1-s}\omega^s\Theta(\omega_c-\omega)$ and  $0<s<1$: &\\  

\parbox{8.5cm}{
for $\omega_c^{-1} \ll t \ll T^{-1}$
\quad
$|\langle\sigma_+(t)\rangle| \approx e^{-\alpha(\omega_0 \,t)^{1-s}}$\\
for $\ {\phantom{\omega_c^{-1} \ll}}t \gg T^{-1}$
\quad
$|\langle\sigma_+(t)\rangle| \approx e^{-\alpha T\,t\,(\omega_0 \,t)^{1-s}}$
}
&
\parbox{5cm}{
\psfig{width=4cm,figure=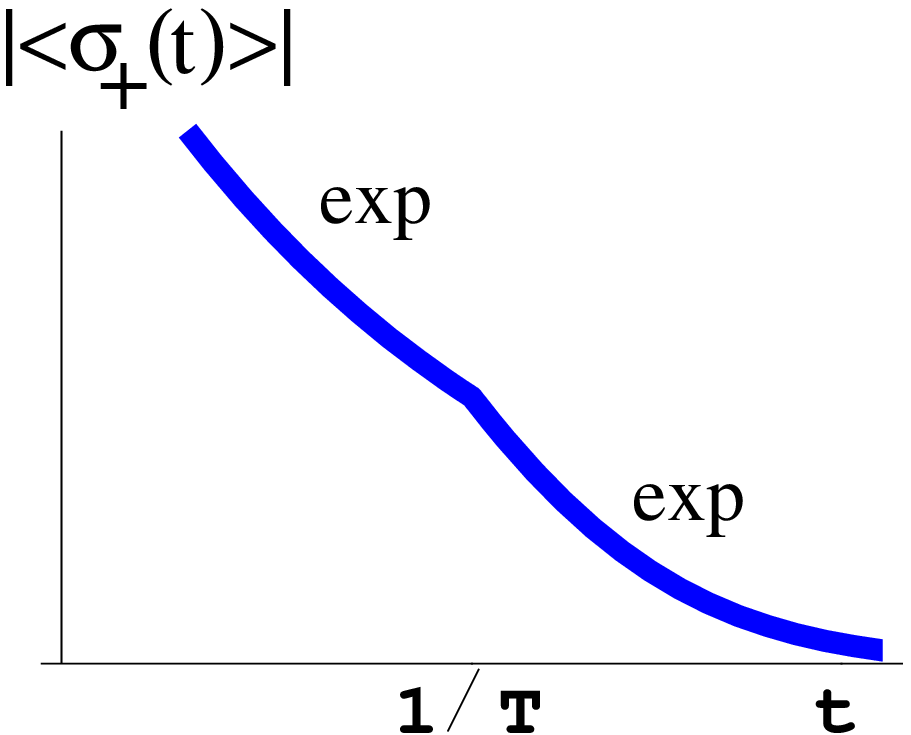}
}\\

\noindent 
$\bullet$
Super-Ohmic: $J(\omega) = \frac{\pi}{2}\alpha \omega_0^{1-s}\omega^s
\Theta(\omega_c-\omega)$ and $1<s<2$:&\\ 

\parbox{8.5cm}{
for $\omega_c^{-1} \ll t \ll T^{-1}$ \quad
$|\langle\sigma_+(t)\rangle| \approx e^{-\alpha(\omega_c/\omega_0)^{s-1}}$\\
for $\ {\phantom{\omega_c^{-1} \ll}}t \gg T^{-1}$ \quad
$|\langle\sigma_+(t)\rangle| \approx e^{-\alpha T\,t\,(\omega_0 \,t)^{1-s}}$
}
&
\parbox{5cm}{
\psfig{width=4cm,figure=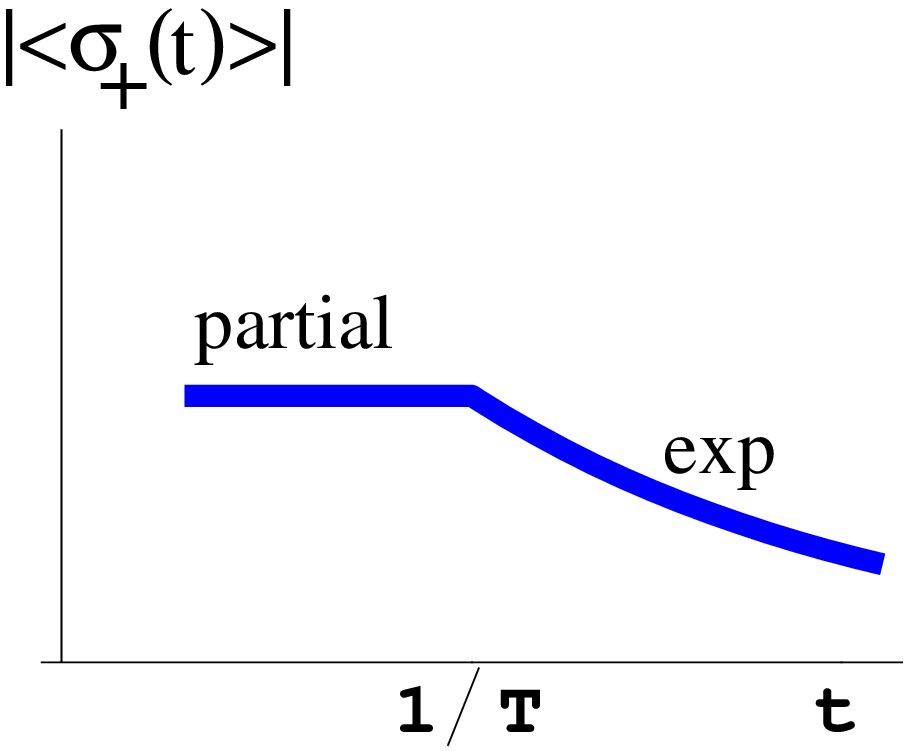}
}\\

\noindent 
$\bullet$
Soft gap: $J(\omega) = \frac{\pi}{2}\alpha
\omega_0^{1-s}\omega^s\Theta(\omega_c-\omega)$ and $s>2$:&\\ 
\parbox{8.5cm}{
for $\omega_c^{-1} \ll t \ \phantom{ \ll T^{-1}}$ \quad
$|\langle\sigma_+(t)\rangle| \approx e^{-\alpha(\omega_c/\omega_0)^{s-1}}$
}
&
\parbox{5cm}{
\psfig{width=4cm,figure=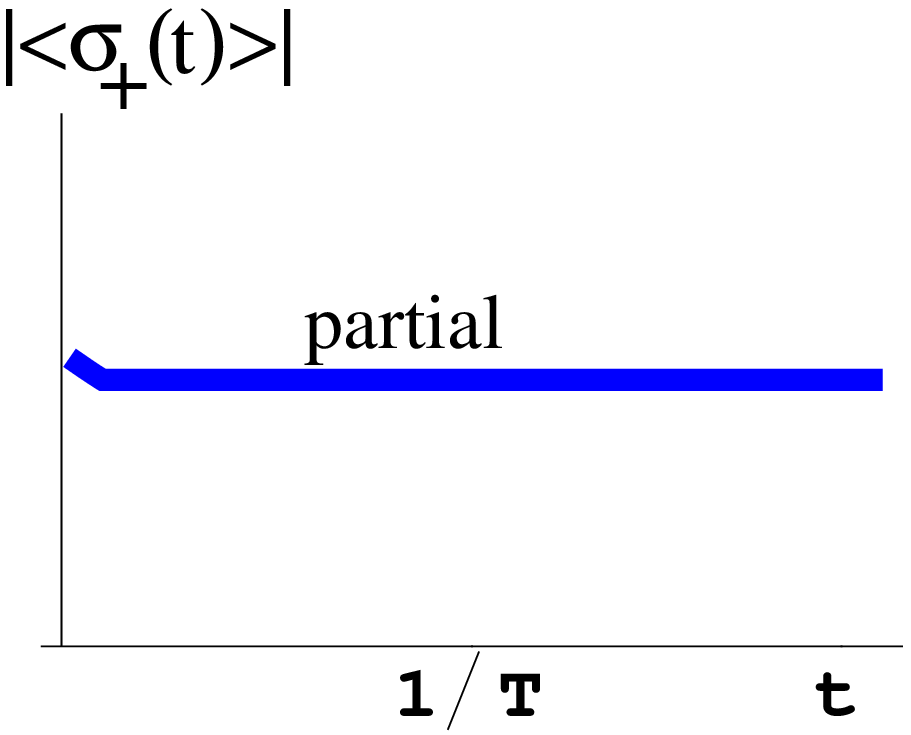}
}

\end{tabular}
\end{table}

\section{Preparation effects}
\label{sec:preparation}

In the previous section we considered specific initial conditions with
a factorized density matrix.
The bath was prepared in an equilibrium state characterized by temperature 
$T$, while the spin state was arbitrary. The state of the total system 
is, thus, a nonequilibrium one and dephasing  is to be expected
even at zero bath temperature. 
In this section we investigate what initial conditions may arise 
in real experiments. 

The fully factorized initial state just described can in principle be 
prepared by the following Gedanken experiment:
The spin is forced, e.g., by a strong external field,
to be in a fixed state, say $\ket{\uparrow}$. The bath, which is kept   
coupled to the spin, relaxes to the equilibrium state of the 
Hamiltonian ${\cal H}_{\uparrow}$, e.g., at $T=0$ to the ground state
$\ket{{\rm g}_{\uparrow}}$ of ${\cal H}_{\uparrow}$.  
Then, at $t=0$, a 
{\sl sudden} pulse of the external field is applied to change 
the spin state, e.g., to a superposition 
$\frac{1}{\sqrt{2}}(\ket{\uparrow} + \ket{\downarrow})$. 
Since the bath has no time to respond, the resulting state is
$\ket{\rm i} = \frac{1}{\sqrt{2}}
(\ket{\uparrow} + \ket{\downarrow}) \otimes \ket{{\rm
g}_{\uparrow}}$.
Both components of this initial wave function now evolve in
time according to the Hamiltonian (\ref{Eq:SpinBoson}). 
The first, $\ket{\uparrow} \otimes \ket{{\rm
g}_{\uparrow}}$, which is an eigenstate of
(\ref{Eq:SpinBoson}), acquires only a trivial phase factor.  The time
evolution of the second component is more involved.  Up to a phase
factor it is given by $\ket{\downarrow} \otimes \exp(-i{\cal
H}_{\downarrow}t/\hbar) \ket{{\rm g}_{\uparrow}}$ where ${\cal
H}_{\downarrow}\equiv {\cal H}_{\rm b}-\sum_j c_j 
(a^{\phantom\dagger}_j+a^\dagger_j)$.  As the
state $\ket{{\rm g}_{\uparrow}}$ is not an eigenstate of 
${\cal H}_{\downarrow}$, entanglement between the spin and the
bath develops, and the coherence between the components of the spin's
state is reduced by 
the factor $|\,\bra{{\rm g}_{\uparrow}} \exp(-i{\cal
H}_{\downarrow}t/\hbar) \ket{{\rm g}_{\uparrow}}\,| =
|\,\bra{{\rm g}_{0}} \exp(-i\Phi(t)) \exp(i\Phi) \ket{{\rm g}_{0}}\,|=
|\,P_{\omega_c}(t,T=0)| < 1$. The function $P(t)$ was defined 
in Eq.~(\ref{Eq:P(t)}). The subscript $\omega_c$ is added to  
indicate the value of the high-frequency cut-off of the bath which  
will play an important role in what follows.   

In a real experiment of the type discussed the preparation pulse 
takes a finite time, $\tau_p$. 
For instance, the $(\pi/2)_x$-pulse 
which transforms the state $\ket{\uparrow}\to 
\frac{1}{\sqrt{2}} (\ket{\uparrow}+\ket{\downarrow})$, can be accomplished by
putting $B_z=0$ and $B_x=\hbar\omega_p$ for a time span
$\tau_p=\pi/2\omega_p$. 
During this time the bath oscillators partially adjust to the  
changing spin state. The oscillators 
with (high) frequencies, $\omega_j \gg \omega_p$, 
follow the spin adiabatically. 
In contrast, the oscillators with low frequency, 
$\omega_j\ll\omega_p$, do not change their state. 
Assuming that the oscillators can be split into these two 
groups, we see that just after the 
$(\pi/2)_x$-pulse the state of the system is 
$\frac{1}{\sqrt{2}}\left(\ket{\uparrow}\otimes\ket{{\rm
g}_{\uparrow}^{\rm h}} + \ket{\downarrow}\otimes\ket{{\rm
g}_{\downarrow}^{\rm h}}\right) \otimes \ket{{\rm g}_{\uparrow}^{\rm
l}}$ where the superscripts `h' and `l' refer to high-
and low-frequency oscillators, respectively.
Thus, we arrive at an initial state with only the low-frequency 
oscillators factorized from the spin.
For the off-diagonal element of the density matrix we obtain
\begin{equation}
|\langle \sigma_+(t) \rangle| = Z(\omega_c,\omega_p)|P_{\omega_p}(t)|
 \, ,
\label{ZP}
\end{equation}
where 
$Z(\omega_c,\omega_p) \equiv 
|\langle g_{\uparrow}^{\rm h} | g_{\downarrow}^{\rm h}\rangle|$
and $P_{\omega_p}(t)$ is given by the same expressions as before,
except that the high-frequency cut-off is reduced to $\omega_p$.

The high frequency oscillators still contribute to the 
reduction of $|\langle \sigma_+(t) \rangle|$ -- via the factor
$Z(\omega_c,\omega_p)$ --  however, 
this effect is reversible. To illustrate this we consider the following 
continuation of the experiment. After the $(\pi/2)$ pulse
we allow for a free evolution of the system during time $t$
with magnetic field $B_z = \Delta E$ along $z$ axis. Then 
we apply a $(-\pi/2)$ pulse and measure $\sigma_z$. Without 
dissipation the result would be $\langle\sigma_z\rangle = \cos(\Delta E t)$.
With dissipation the state of the system after time $t$ is
\begin{equation}
\frac{1}{\sqrt{2}}\left(e^{i\Delta E t/2}\ket{\uparrow}\otimes\ket{{\rm
g}_{\uparrow}^{\rm h}}\otimes \ket{{\rm g}_{\uparrow}^{\rm
l}} + e^{-i\Delta E t/2}\ket{\downarrow}\otimes\ket{{\rm
g}_{\downarrow}^{\rm h}}\otimes e^{-i{\cal
H}_{\downarrow}t/\hbar}\ket{{\rm g}_{\uparrow}^{\rm
l}}\right)
\ .
\end{equation}
After the $(-\pi/2)$ pulse (also of width $\pi/2\omega_p$) 
we obtain the following state:
\begin{eqnarray}
&&\frac{1}{2} \ket{\uparrow}\otimes\ket{{\rm
g}_{\uparrow}^{\rm h}} \otimes
\left(
e^{i\Delta E t/2} \ket{{\rm g}_{\uparrow}^{\rm l}} + 
e^{-i\Delta E t/2}e^{-i{\cal
H}_{\downarrow}t/\hbar}\ket{{\rm g}_{\uparrow}^{\rm l}}
\right)+
\nonumber \\
&&\frac{1}{2} \ket{\downarrow}\otimes\ket{{\rm
g}_{\downarrow}^{\rm h}} \otimes
\left(-\,
e^{i\Delta E t/2} \ket{{\rm g}_{\uparrow}^{\rm l}} + 
e^{-i\Delta E t/2}e^{-i{\cal H}_{\downarrow}t/\hbar}\ket{{\rm
g}_{\uparrow}^{\rm l}} 
\right)
\ .
\end{eqnarray} 
From this we finally get $\langle\sigma_z\rangle = 
{\rm Re} \left[P_{\omega_p}(t) e^{-i\Delta E t}\right]$.
Thus the amplitude of the coherent oscillations of $\sigma_z$ 
is reduced only by the factor $|P_{\omega_p}(t)|$ associated with 
slow oscillators. The high frequency factor $Z(\omega_c,\omega_p)$ 
does not appear. To interpret this result we
note that we could have discussed the experiment in terms of
renormalized spins $|\tilde\uparrow\rangle \equiv 
|\uparrow\rangle|g^{\rm h}_\uparrow \rangle$ and
$|\tilde\downarrow\rangle \equiv 
|\downarrow\rangle|g^{\rm h}_\downarrow \rangle$, and assuming that 
the high frequency cutoff of the bath is $\omega_p$. 

It is interesting to compare further the two scenarios with instantaneous and
finite-time preparation further. The time evolution after the
instantaneous preparation is governed by $P_{\omega_c}(t)$. For 
$T\ll\omega_p$ and $t \gg 1/\omega_p$
and arbitrary spectral density $J(\omega)$
it satisfies the following relation:
$P_{\omega_c}(t) = Z^2(\omega_c,\omega_p)P_{\omega_p}(t)$,
which follows from $\,\bra{{\rm g}^{\rm h}_{\uparrow}} \exp(-i{\cal
H}_{\downarrow}t/\hbar) \ket{{\rm g}^{\rm h}_{\uparrow}} 
\rightarrow |\langle g_{\uparrow}^{\rm h} | g_{\downarrow}^{\rm h}\rangle|^2$
for $t \gg 1/\omega_p$. 
Thus for the instantaneous preparation  the reduction due to the high frequency 
oscillators is equal to $Z^2(\omega_c,\omega_p)$, while a look at
the finite-time preparation result (\ref{ZP}) shows 
that in this case the reduction is weaker, given by a single
power of $Z(\omega_c,\omega_p)$ only. 
Moreover, in the slow preparation experiment the factor $Z(\omega_c,\omega_p)$
originates from the overlap of two `simple' wave functions, 
$\,\ket{{\rm g}^{\rm h}_{\uparrow}}$ and 
$\,\ket{{\rm g}^{\rm h}_{\downarrow}}$, which can be further 
adiabatically manipulated, as described  above, and this
reduction can be recovered. This effect is to be interpreted as 
a renormalization.   
On the other hand, for the instantaneous preparation the high frequency 
contribution to the dephasing originates from the overlap of 
the states $\,\ket{{\rm g}^{\rm h}_{\uparrow}}$and  $e^{-i{\cal
H}_{\downarrow}t/\hbar}\ket{{\rm g}_{\uparrow}^{\rm h}}$. The latter is a 
complicated excited state of the bath with many nonzero amplitudes 
evolving with different frequencies. There is no simple (macroscopic) way
to reverse the dephasing associated with this state. 
Thus we observe that the time scale of the manipulating pulses 
determines the border between the oscillators responsible for 
dephasing and the oscillators responsible for renormalization.

\section{Response functions}

In the limit $\theta = 0$ we can also calculate exactly the
linear response of $\tau_x=\sigma_x$ to a weak magnetic field in
the $x$-direction, $B_x(t)$:
\begin{equation}
\label{Eq:response function}
\chi_{xx}(t) = \frac{i}{\hbar}\;\Theta(t) \langle
\tau_x(t)\tau_x(0)-\tau_x(0)\tau_x(t)\rangle \ .
\end{equation}  
Using the equilibrium density matrix
\begin{equation}
\hat \rho^{\rm eq} = (1+e^{-\beta\Delta E})^{-1}
\left[\ket{\uparrow}\bra{\uparrow}\otimes \hat\rho_{\uparrow} +
e^{-\beta\Delta E} \ket{\downarrow}\bra{\downarrow}\otimes
\hat\rho_{\downarrow} \right]
\ ,
\end{equation}
where $\hat\rho_{\uparrow} \propto \exp(-\beta {\cal H}_{\uparrow})$ is the
bath density matrix adjusted to the spin state $\ket{\uparrow}$, and
similar for $\hat\rho_{\downarrow}$, 
we obtain the susceptibility
\begin{equation}
\chi_{xx}(t) = -\frac{2\hbar^{-1}\Theta(t)}{1+e^{-\beta\Delta E}}\;{\rm Im}
\left[P_{\omega_c}(t)e^{-i\Delta E t} + 
e^{-\beta \Delta E}P_{\omega_c}(t)e^{i\Delta E t}\right] \ .
\end{equation} 
Thus, to calculate the response function one has to use the full
factor $P_{\omega_c}(t)$, corresponding to the instantaneous 
preparation of an initial state. This can be understood by looking at 
the Kubo formula (\ref{Eq:response function}). The operator $\tau_x(0)$ 
flips only the bare spin without touching the oscillators, as if an infinitely 
sharp $(\pi/2)$ pulse was applied.

The imaginary part of the Fourier transform of $\chi(t)$, which
describes dissipation,  is
\begin{equation}
\chi_{xx}''(\omega) = \frac{1}{2(1+e^{-\beta \Delta E})}
\left[P(\hbar\omega-\Delta E)+ e^{-\beta \Delta E} P(\hbar\omega+\Delta E)
\right] - ...(-\omega) \ .
\end{equation}
At $T=0$ and positive values of $\omega$ we use
the expression for $P(E)$ from Ref.~\cite{P(E)_Devoret}
to obtain
\begin{equation}
\chi_{xx}''(\omega) = \frac{1}{2}P(\hbar\omega-\Delta E)=
\Theta(\hbar\omega-\Delta E)
\frac{e^{-2\gamma\alpha}(\hbar\omega_c)^{-2\alpha}}{2\Gamma(2\alpha)}
(\hbar\omega-\Delta E)^{2\alpha -1} \ .
\end{equation}
We observe that the dissipative part
$\chi_{xx}''$ has a gap $\Delta E$, which corresponds to 
the minimal energy needed to flip the spin, 
and a power-law behavior  as $\omega$
approaches the threshold. This behavior of $\chi_{xx}''(\omega)$
is known from the {\it orthogonality catastrophe }
scenario~\cite{Mahan}. It implies that the ground state of the
oscillator bath for  different spin states,
$\ket{{\rm g}_{\uparrow}}$ and $\ket{{\rm g}_{\downarrow}}$, are
{\it macroscopically} orthogonal. In particular, for an Ohmic bath 
we recover the behavior typical for the problem of 
X-ray absorption in metals~\cite{Mahan}.

As $\chi''(\omega)$ characterizes the dissipation in the system 
(absorption of energy from the perturbing magnetic field) it is 
interesting to understand the respective roles of high- and 
low-frequency oscillators. We use the spectral decomposition 
for $\chi''$ at $T=0$,
\begin{equation}
\chi''(\omega) = \pi \sum_n |\bra{0} \tau_x \ket{n}|^2
\left[\delta(\omega-E_n) - \delta(\omega+E_n)\right]
\ ,
\end{equation} 
where $n$ denotes exact eigenstates of the system. These are 
$\ket{\uparrow} \ket{n_{\uparrow}}$ and 
$\ket{\downarrow} \ket{n_{\downarrow}}$, where 
$\ket{n_{\uparrow}}$ and $\ket{n_{\uparrow}}$ denote the excited
(multi-oscillator) states 
of the Hamiltonians ${\cal H}_{\uparrow}$ and ${\cal H}_{\downarrow}$.
The ground state is $\ket{\uparrow} \ket{g_{\uparrow}}$ and the 
only excited states that contribute to $\chi''(\omega)$ 
are $\ket{\downarrow} \ket{n_{\downarrow}}$ with 
${\cal H}_{\downarrow}\ket{n_{\downarrow}} = 
(\omega - \Delta E)\ket{n_{\downarrow}}$. This means that all 
the oscillators with frequencies $\omega_j > \omega - \Delta E$
have to be in the ground state. Therefore we obtain for $\omega_p >
\omega - \Delta E$ 
\begin{equation}
\chi_{\omega_c}''(\omega) = 
Z^2(\omega_c,\omega_p) \chi_{\omega_p}''(\omega) \ .
\label{Eq:renormalization_of_chi}
\end{equation} 

To interpret this result we generalize 
the coupling to the magnetic field by introducing a $g$ factor:
$H_{\rm int}=-(g/2) \delta B_x(t) \sigma_x$. Then, if the applied 
magnetic field can be independently measured, the observable quantity 
corresponding, e.g., to the energy absorption is 
\begin{equation}
\label{Eq:g_response function}
\chi_{xx}(t) = g^2\frac{i}{\hbar}\;\Theta(t) \langle
\left[\sigma_x(t),\sigma_x(0)\right]\rangle \ .
\end{equation}  
Thus, Eq.~(\ref{Eq:renormalization_of_chi}) tells us that 
by measuring the response of the spin at frequencies 
$\omega < \omega_p + \Delta E$ we cannot distinguish 
between a model with upper cutoff $\omega_c$ and 
$g=1$ and a model with cutoff $\omega_p$ 
and $g=Z(\omega_{\rm c},\omega_{\rm p})$. This is the 
usual situation in the renormalization group context. 
Thus, again, we note that high-frequency oscillators are naturally 
associated with renormalization effects.  \\

We collect the results for $\chi''$ for various spectra 
in Table~\ref{tbl:chi}. The results are shown for temperatures 
lower and higher than the crossover temperature introduced in 
Section~\ref{sec:long_exact}.\\

\begin{table}
\caption{\label{tbl:chi}Response functions for different bath spectra.}
\begin{tabular}{l|l}
&\\
\parbox{3.0cm}{
Bath spectrum \\
$J(\omega) =\frac{\pi}{2} \alpha\omega_0^{1-s}\omega^s$\\
$\times \Theta(\omega_c-\omega)$
}
&
\parbox{7.5cm}{
Response function\\
$\chi_{xx}(t) = i\Theta(t) \langle
[\sigma_x(t),\sigma_x(0)]\rangle$
\\[0.2cm]
\parbox{3.7cm}{$\chi''_{xx}(\omega)$, $T<T^*$}
\parbox{3.7cm}{$\chi_{xx}''(\omega)$, $T>T^*$}
}
\\ 
&\\
\hline
&\\
\parbox{3.0cm}{
Ohmic \\ $s=1$
} 
&
\parbox{7.5cm}{
\parbox{3.7cm}{\hskip -0.cm
\psfig{width=3.7cm,figure=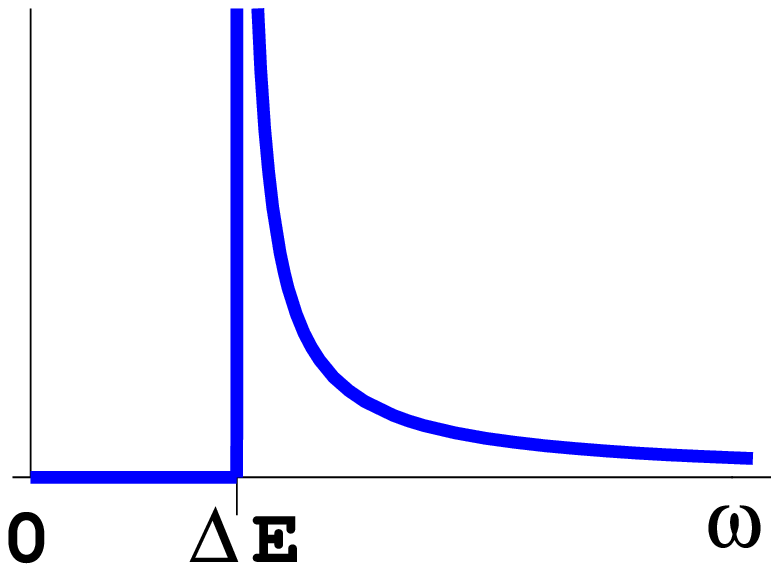}
$\frac{\alpha\,\Theta(\omega-\Delta E)}{(\omega-\Delta E)^{(1-2\alpha)}}$
}
\parbox{3.7cm}{\hskip -0.cm
\psfig{width=3.7cm,figure=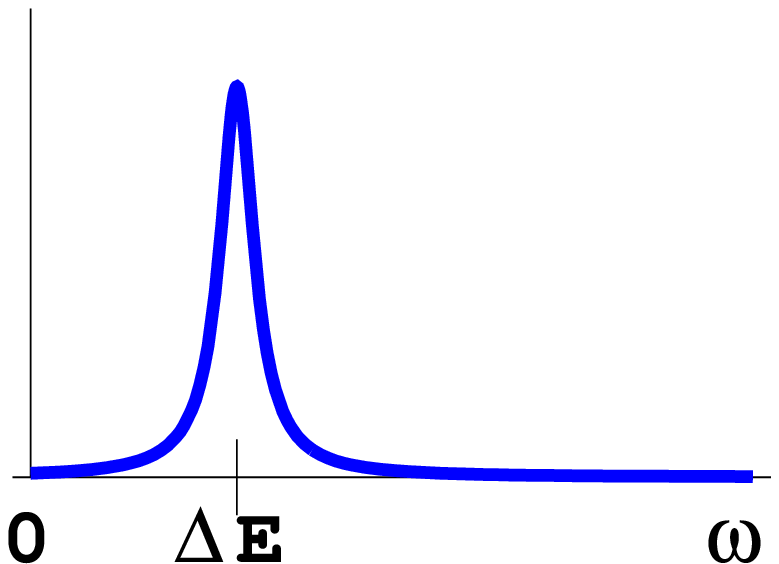}
$\frac{\alpha\,T}{(\alpha T)^2+(\omega-\Delta E)^2}$
}
}
\\
&\\
\hline
&\\
\parbox{3.0cm}{
Sub-Ohmic \\ $0 \le s<1$
}  
&
\parbox{7.5cm}{
\parbox{3.7cm}{\hskip -0.cm
\psfig{width=3.7cm,figure=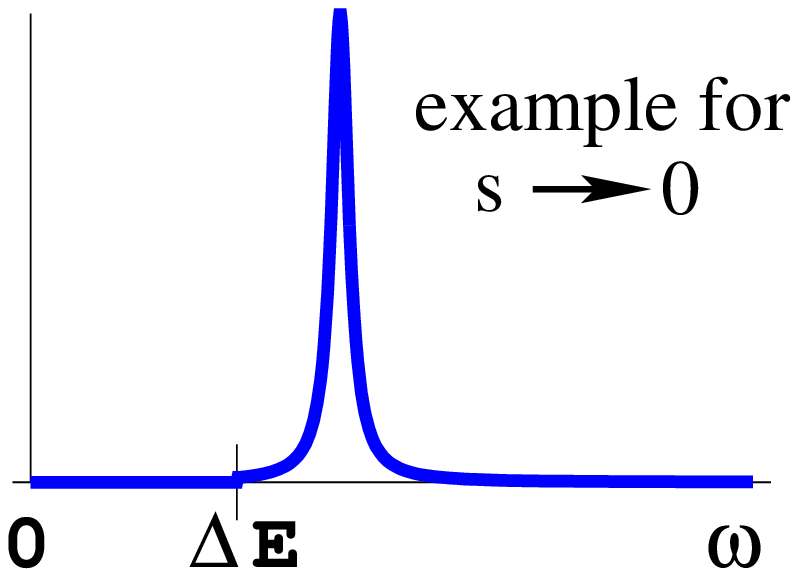}
$\frac{\alpha\,\omega_0\,\theta(\omega-\Delta E)}
{(\alpha \omega_0)^2+
(\omega-\Delta E-\alpha\omega_0\ln(...))^2}$
}
\parbox{3.7cm}{\hskip -0.cm
\psfig{width=3.7cm,figure=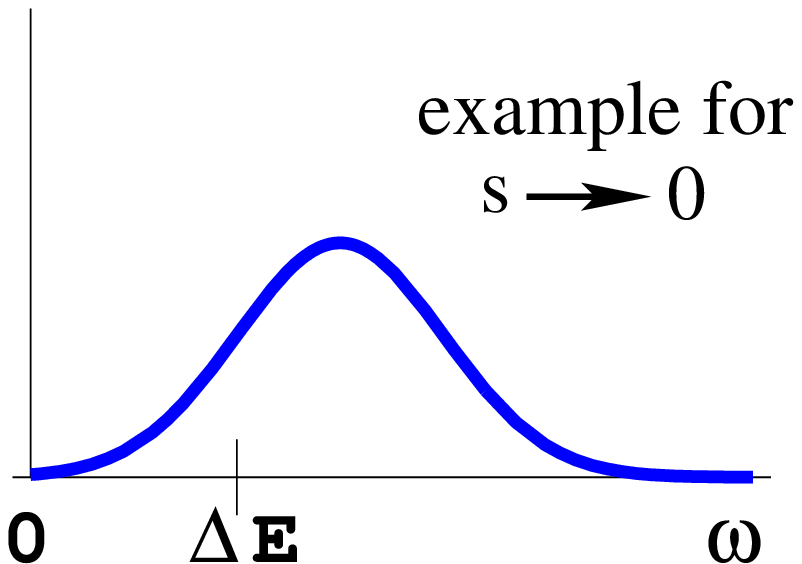}
$\exp\left[-\frac{(\omega-\Delta E-\alpha\omega_0\ln(...))^2}
{\alpha\omega_0 T\ln(...)}\right]$
}
}
\\
&\\
\hline
&\\
\parbox{3.2cm}{
Super-Ohmic \\ $1 < s\le 2$
}
&  
\parbox{7.5cm}{
\parbox{3.7cm}{\hskip -0.cm
\psfig{width=3.7cm,figure=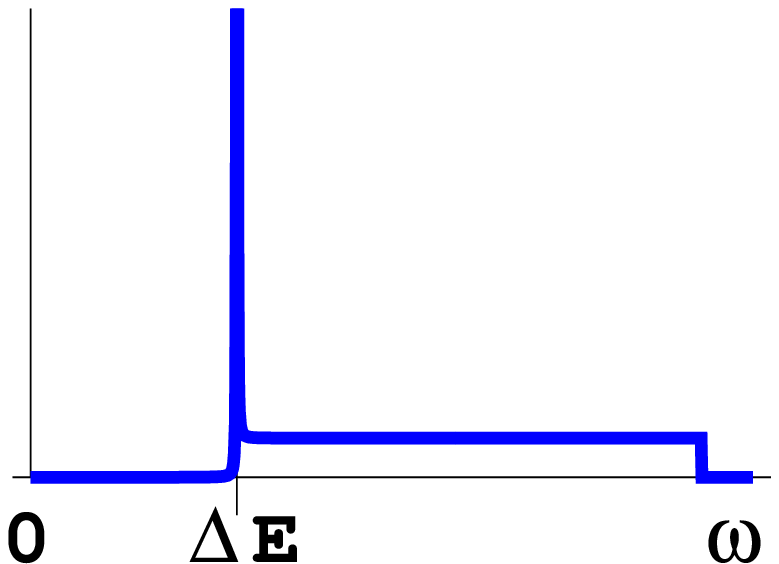}\\
$Z^2\,\delta(\omega-\Delta E)+...$
}
\parbox{3.7cm}{\hskip -0.cm
\psfig{width=3.7cm,figure=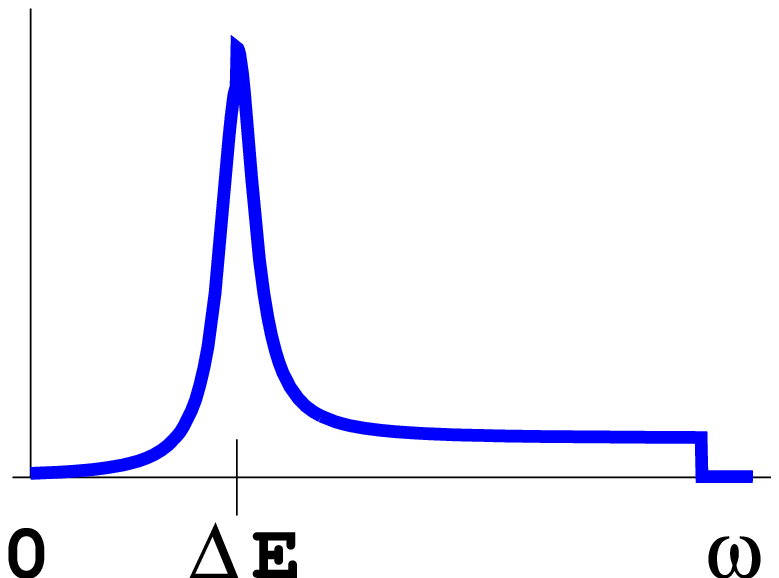}
$Z^2\,\delta_{\Gamma_\varphi}(\omega-\Delta E)+...$
}
}
\end{tabular}
\end{table}

\section{Summary}

The examples presented above show that  
for a quantum two-state system with a non-equilibrium initial 
state, described by a factorized initial density matrix,
dephasing persists down to zero bath temperature. 
An Ohmic environment leads to a power-law dephasing at $T=0$, while a
sub-Ohmic bath yields exponential dephasing. The reason is that 
the factorized initial state, even with the bath in the ground state
of the bath Hamiltonian, is actually a
superposition of many excited states of the total coupled system. 
In a real experiment only a part of the 
environment, the oscillators with low frequencies, can be prepared
factorized from the two level system. These oscillators still lead to
dephasing, whereas the high-frequency oscillators lead to
renormalization effects. 
The examples demonstrate that experimental conditions, e.g., 
details of the system's initial state preparation, 
determine which part of the environment contributes to dephasing 
and which part leads to renormalization.
The finite preparation time 
$\sim 1/\omega_p$ also introduces a natural high-frequency cutoff in
the description of dephasing effects.
We have further demonstrated that dephasing and renormalization effects 
influence the response functions of the two level system. 
We noted that they exhibit features known for the orthogonality catastrophe, 
including a power-law divergence above a threshold.

\section{Acknowledgments}

We thank Y. Makhlin for valuable contributions to the present work
and M.~B\"uttiker, M.H.~Devoret, D.~Esteve, 
Y.~Gefen, D.~Golubev, Y.~Imry, D.~Loss, A.D.~Mirlin, 
A.~Rosch, U.~Weiss, R.~Whitney, and A.D.~Zaikin
for stimulating discussions. The work is part of the EU IST 
Project SQUBIT and of the {\bf CFN} (Center for Functional Nanostructures) 
which is supported by the DFG (German Science Foundation).

\bibliography{ref}  

\end{document}